# Modeling of the evolution of dielectric loss with processing temperature in ferroelectric and dielectric thin oxide films

X. H. Zhu,[1] B. Guigues,[1-3] E. Defaÿ,[1,a] and M. Aïd[1]

[1]*CEA-LETI MINATEC, 17 Rue des Martyrs, 38054 Grenoble Cedex 9, France*

[2]*Ecole Centrale Paris, Grande Voie des Vignes, 92295 Châtenay Malabry Cedex, France*

[3]*ST Microelectronics, 850 av Jean Monnet, F-38926 Crolles Cedex, France*

**Abstract**

It was experimentally found that the evolution of dielectric loss with processing temperature displays a common trend in ferroelectric and dielectric thin oxide films: firstly an increase and then a decrease in dielectric loss when the processing temperature is gradually raised. Such a dielectric response of ferroelectric/dielectric thin films has been theoretically addressed in this work. We propose that at the initial stage of the crystallization process in thin films, the transformation from amorphous to crystalline phase should increase substantially the dielectric loss; then, with further increase in the processing temperature, the coalescent growth of small crystalline grains into big ones could be helpful in reducing the dielectric loss by lowering grain boundary densities. The obtained experimental data for $(Ba,Sr)TiO_3$ thin films with 500 nm in thickness were analyzed in terms of the model developed and shown to be in a reasonable agreement with the theoretical results.

[a]Author to whom correspondence should be addressed; electronic mail: edefay@cea.fr

# I. Introduction

In recent years, ferroelectric and dielectric thin oxide films have attracted considerable attention for applications in various electronic devices, such as high-value capacitors, electrically tunable microwave devices, and random access memories, etc [1-5]. For these applications, one of key issues is to understand their loss mechanisms and to adjust effectively the dielectric loss in order to meet the requirements of real devices. It is known that there are two different classes of contributions to dielectric loss: (i) intrinsic loss due to the interaction of the ac field with the phonons of the material and (ii) extrinsic loss owing to the coupling of the microwave field with defects [5-9]. According to the theory of quantum mechanics, the intrinsic loss mainly corresponds to the absorption of the energetic quantum of the electromagnetic field $\hbar\omega$ ($\omega$ is the ac field frequency) in collisions with the thermal phonons. The role of the fundamental phonon loss mechanisms in the total balance of the dielectric loss of a material is believed to be strongly dependent on the dielectric permittivity of the material and the measuring frequency: typically, the higher the frequency and permittivity, the more important the intrinsic loss [5]. On the other hand, the extrinsic dielectric loss in ferroelectric/dielectric thin films could stem from charged defects and local polar regions. Charged defects are usually associated with oxygen vacancies within nonstoichiometric grain boundaries, whereas the local polar regions are associated with the planar (grain boundaries), linear (dislocations), and/or point (vacancies and/or substituted atoms) defects [10]. It is noteworthy that at relatively high frequencies and/or under the action of a dc bias field, the intrinsic and extrinsic contributions could be comparable so that the dominant contribution to the loss may be extrinsic or intrinsic relying on the quality of the material. Because the preparation procedure can affect greatly the sample quality and, thus, physical properties, the identification of the dominant loss mechanisms and their correlation with the process parameters may help to develop methods of improving the ferroelectric/dielectric film microstructure and thereby reduce the dielectric loss down to the fundamental limit.

In this work, the evolution of dielectric loss with processing temperature is experimentally investigated and theoretically analyzed for ferroelectric (Ba,Sr)TiO$_3$ (BST) thin films and pyrochlore-type Pb(Mg,Nb,Ti)O$_3$ (PMNT) dielectric films. The modeling results provide important practical implications for tailoring the dielectric loss in ferroelectric and dielectric thin films.

## II. Experimental

Thin films of BST and PMNT were prepared on technologically desirable Pt/TiO$_2$/SiO$_2$/Si(100) substrates by physical vapor deposition method. Ba$_{0.7}$Sr$_{0.3}$TiO$_3$ films, with two different thicknesses of 100 and 500 nm, were ion beam sputtered from a single stoichiometric target. PMNT thin films with thickness of 50 nm were sputtered from a ceramic target with a nominal composition of 0.9Pb(Mg$_{1/3}$Nb$_{2/3}$)O$_3$-0.1PbTiO$_3$ and an excess of 15 mol% MgO. The substrates were not heated during the deposition and their temperature was only dependent on ion bombardment. After deposition, the films were post-annealed at a wide range of temperatures, ranging from 300 to 750 °C, for 30 min in a conventional furnace in air ambient.

The structural and dielectric properties of the BST and PMNT films were characterized. Their crystallographic structures were analyzed by x-ray diffraction (XRD) with a Siemens D5000 diffractometer using CuK$_\alpha$ radiation. The film microstructure and crystalline grain growth were checked by atomic force microscopy (AFM) using a tapping mode (Digital Instruments/Veeco Dimension 3000). In order to measure the dielectric properties of the films, circular top platinum electrodes with 110 μm in diameter were patterned by photolithography and lift-off process. The dielectric property measurements were performed at room temperature by using an HP4194A impedance analyzer or an HP4275A multi-frequency LCR meter.

**III. Method for modeling of dielectric loss**

In the initial stage of crystallization, there is a transition from amorphous to crystalline phase with the increase in the processing temperature. This can be considered as crystalline spheres being embedded into an amorphous material matrix, which has been evidenced by transmission electron microscopy observations [11]. The spherical inclusion model is schematically illustrated in figure 1. It is proposed that a modified effective medium approximation (MEMA) approach could be applied to describe the dielectric responses of ferroelectric-dielectric composites [12]. Accordingly, the effective loss tangent of the composite can be given by

$$\tan \delta_{comp} = \frac{1}{\varepsilon_{comp}} \cdot \frac{(1-q)c_a^2 \varepsilon_a \tan \delta_a + q c_c^2 \varepsilon_c \tan \delta_c}{1 - 2(1-q)b_a^2 - 2q b_c^2}, \quad (1)$$

where $q$ is the volume concentration of crystallites, $\varepsilon_a$, $\varepsilon_c$, $\tan\delta_a$, and $\tan\delta_c$ are, respectively, the typical dielectric constant and loss tangent for amorphous phase and fully crystallized phase, the coefficients of $c_a$, $c_c$, $b_a$, and $b_c$ can be expressed as

$$c_a = 3\varepsilon_{comp}/(\varepsilon_a + 2\varepsilon_{comp}), \quad (2)$$

$$c_c = 3\varepsilon_{comp}/(\varepsilon_c + 2\varepsilon_{comp}), \quad (3)$$

$$b_a = (\varepsilon_a - \varepsilon_{comp})/(\varepsilon_a + 2\varepsilon_{comp}), \quad (4)$$

$$b_c = (\varepsilon_c - \varepsilon_{comp})/(\varepsilon_c + 2\varepsilon_{comp}), \quad (5)$$

and $\varepsilon_{comp}$ is the effective permittivity of the composite, which takes the form

$$\varepsilon_{comp} = \frac{1}{4}\left\{-\varepsilon_a + 3(1-q)\varepsilon_a + 2\varepsilon_c - 3(1-q)\varepsilon_c + \sqrt{8\varepsilon_a \varepsilon_c + [-\varepsilon_a + 3(1-q)\varepsilon_a + 2\varepsilon_c - 3(1-q)\varepsilon_c]^2}\right\}. \quad (6)$$

Based on this model, it is expected that with the improvement in crystallization, dielectric loss should be increased along with enhanced dielectric permittivity.

During the crystallization, a large number of crystalline grains are generated. This leads to a more pronounced extrinsic loss contribution induced by grain boundaries. However, when the material is sufficiently crystallized, the coalescent growth of small crystalline grains into big ones could reduce the dielectric loss by lowering the density of grain boundaries. Due to the change of the average coordination number and bond length of the atoms (or ions) in the grain boundary core, there is an excess free energy which leads to depletion or accumulation effects for the mobile ionic and/or electronic particles, thereby leading to resistance degradation [13]. According to the results reported by Tagantsev et al. [5], the dielectric loss due to local polar regions can be evaluated by the "defect-induced" quasi-Debye mechanism:

$$\tan \delta_{QD} \propto \varepsilon^{4.5-d}, \tag{7}$$

where $d$ is the dimension of the defect, i.e., $d=2$ for a planar defect like grain boundaries, $d=1$ for linear defects, and $d=0$ for point defects. Hence, the dielectric loss caused by grain boundaries can be approximated as follows:

$$\tan \delta_{GB} = C \cdot S \cdot \varepsilon^{2.5}, \tag{8}$$

where $C$ is a structure-related constant coefficient and $S$ is the grain boundary density. For the simplicity of measurement and calculation, we transform the density ($S$) from $a \dfrac{\sum_i 4\pi r_i^2}{V}$, the total surface area of grain boundaries within a given volume, to $b \dfrac{\sum_i 2\pi r_i}{A}$, the total perimeter for the projection of grain boundaries over a given area. Accordingly, it is quite convenient to derive the evolution of grain boundary density during the crystallization process, as presented in Fig. 2. On the basis of this simplified simulation, it can be concluded that the density of grain boundaries firstly increases and then decreases with the improvement in material crystallization.

To sum up, the total effective dielectric loss can be written as

$$\tan \delta_{eff} = \tan \delta_{comp} + \tan \delta_{GB}. \tag{9}$$

It should be noted that we only consider the main contributions; the dielectric loss caused by other factors like charged point defects is negligible in our modeling.

**IV. Results and discussion**

Figure 3 shows the evolution of dielectric loss with processing temperature for the BST and PMNT thin films. The dielectric loss measurements were performed at room temperature and at a frequency of 100 kHz with a small ac signal amplitude of 50 mV. It should be pointed out that the BST films are crystallized in perovskite phase while the PMNT films are in pyrochlore phase. Details of crystal structures for them will be published elsewhere. It is interesting to note that these films display a common trend in the evolution of dielectric loss, that is, firstly an increase and then a decrease in dielectric loss when the processing temperature is gradually raised. Owing to the fact that similar dielectric behavior was observed in different kinds of thin oxide films, understanding of the physical mechanisms behind this phenomenon would be of great significance. In the following, we will address this question by taking for example the BST films of 500 nm thickness.

Figure 4 shows $\theta$-$2\theta$ XRD scans of the BST thin films with thickness of 500 nm that were annealed at different temperatures (440, 450, 535, and 685 °C). With the increase in processing temperature, the crystallinity in perovskite phase is gradually improved. For the 440 °C-annealed film, very weak and almost no noticeable diffraction peaks for BST can be detected. This is probably because the crystallized BST nuclei within the amorphous BST matrix are very small [11]. However, when the processing temperature is increased to 450 °C, the perovskite phase can be evidenced; after that, a further increase in processing temperature results in an enhancement of peak intensity. In order to better understand the evolution of crystallinity in the BST films, tapping-mode AFM images were collected, as shown in Fig. 5. Note that the scan area is 1×1 μm$^2$ for all the images. For the 435 °C-annealed film, no crystallites were observed, which is consistent with its amorphous nature revealed by XRD and dielectric property measurements. As the processing temperature is increased above 435 °C, crystallization starts to occur in the BST thin films. As a result, a small amount of crystalline grains appear in the 465 °C-annealed film. Then, both the amount of crystalline grains and their size increase with increasing the processing temperature, indicative of an improvement in the crystallization but also an increased number of grain boundaries. Nonetheless, the small crystalline grains coalesce into bigger ones when the films are sufficiently crystallized, which in turn lowers the density of grain boundaries. By combining the AFM observations with the XRD results, the processing temperature dependence of crystalline volume ratio ($q$) and grain boundary density ($S$) can be estimated, as shown in table I. It is assumed that the distribution of grains is homogeneous across the film thickness in our modeling.

Let us now focus on the theoretical modeling and its comparison to the experimental results. On one hand, by applying the obtained $q$ values as well as the typical dielectric constant and loss tangent of our 500 nm BST films for amorphous phase ($\varepsilon_a$=15 and tan$\delta_a$=0.003) and fully crystallized phase ($\varepsilon_c$=200 and tan$\delta_c$=0.035) to equation (1), the first contribution to dielectric loss by the composite model has been evaluated and plotted with solid triangle symbols in Fig. 6. As expected, an abrupt increase in dielectric loss happens at the amorphous-to-crystalline transition temperature, indicating that the improvement in dielectric polarization is accompanied by higher energy dissipation. Above this transition temperature, the dielectric loss of the composite increases slowly and tends to be stabilized.

On the other hand, the second term of equation (9) can be solved using the obtained $S$ values and an experimentally determined $C$ value ($1.9 \times 10^{-6}$ nm). The calculated data are plotted in Fig. 6 with open circle symbols. With the increase in processing temperature, as mentioned above, the amount of crystalline grains is firstly increased along with the phase transition from amorphous to crystalline, and, thus, the dielectric loss caused by grain boundaries is increased; however, the number of grain boundaries is then decreased because of the coalescent growth of small grains into bigger ones when the processing temperature is relatively high, which in turn reduces significantly the dielectric loss. Consequently, in the initial stage of crystallization, the dielectric loss is dominated by $\tan\delta_{comp}$. For sufficiently crystallized thin films, the $\tan\delta_{comp}$ contribution tends to be stabilized and the dominant mechanism turns to $\tan\delta_{GB}$. As illustrated in Fig. 6, the total effective dielectric loss through modeling is plotted and shown to agree well with the experimental results.

## V. Conclusions

We have experimentally found that the evolution of dielectric loss with processing temperature displays a common trend in perovskite-type ferroelectric BST and pyrochlore-type dielectric PMNT thin films: firstly an increase and then a decrease in dielectric loss when the processing temperature is gradually raised. A theoretical model combining the composite contribution with that of grain boundaries has been proposed in this work to explain the observed dielectric behavior. In the initial stage of crystallization, the transformation from amorphous to crystalline phase increases substantially the dielectric loss due mainly to higher energy dissipation by the enhancement of dielectric polarization; then, with further increase in the processing temperature, the coalescent growth of small crystalline grains into big ones results in the reduction of dielectric loss by lowering grain boundary densities. The obtained experimental data for $Ba_{0.7}Sr_{0.3}TiO_3$ thin films with 500 nm in thickness were analyzed in terms of the model developed and shown to be in a reasonable agreement with the theoretical results. These good modeling results pave the way for seeking methods in order to tailor the dielectric loss effectively for practical applications.

**Table I.** Processing temperature dependent crystalline volume ratio ($q$) and grain boundary density ($S$) for the 500-nm-thick BST films.

| $T$ (°C)       | 435 | 450   | 465   | 485   | 535   | 585   | 635   | 685   |
|----------------|-----|-------|-------|-------|-------|-------|-------|-------|
| $q$ (±5%)      | 0   | 25%   | 35%   | 40%   | 45%   | 60%   | 92%   | 95%   |
| $S$±5% (1/nm)  | 0   | 0.001 | 0.008 | 0.006 | 0.025 | 0.016 | 0.012 | 0.004 |

**Figure captions**

**Figure 1.** Schematic representation of the spherical inclusion model. The dark color corresponds to inclusions of the crystallites.

**Figure 2.** Schematic illustration of the correlation between grain boundary density ($S$) and crystallization process: (a) amorphous matrix ($S=0$), (b) start-up of crystallization with nucleation ($S=24$ units), (c) progress in crystallization with an increase in the number of crystalline grains and grain size ($S=56$ units), and (d) coalescent growth of small grains into big ones ($S=30$ units).

**Figure 3.** Evolution of dielectric loss with processing temperature for the BST and PMNT thin films, measured at 300 K and 100 kHz.

**Figure 4.** XRD $\theta$-$2\theta$ scans of the BST thin films with thickness of 500 nm that were deposited on Pt/TiO$_2$/SiO$_2$/Si substrates and post-annealed at different temperatures.

**Figure 5.** AFM micrographs for the 500-nm-thick BST films annealed at different temperatures: (a) 435 °C, (b) 465 °C, (c) 535 °C, (d) 585 °C, (e) 635 °C, and (f) 685 °C. It is pointed out that the scan area is 1×1 μm$^2$ for all the images.

**Figure 6.** Modeling of dielectric losses as a function of the processing temperature for the BST thin films with 500 nm in thickness, as well as its comparison to the experimental results. Note that the error bars arise from the difficulty in precise estimation of crystalline volume ratio ($q$) and grain boundary density ($S$).

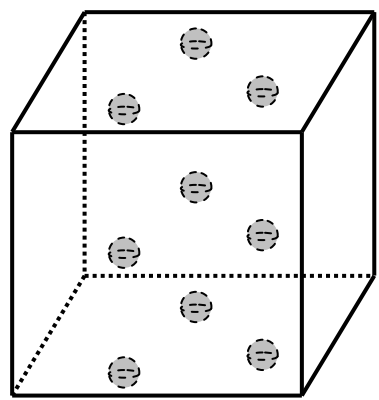

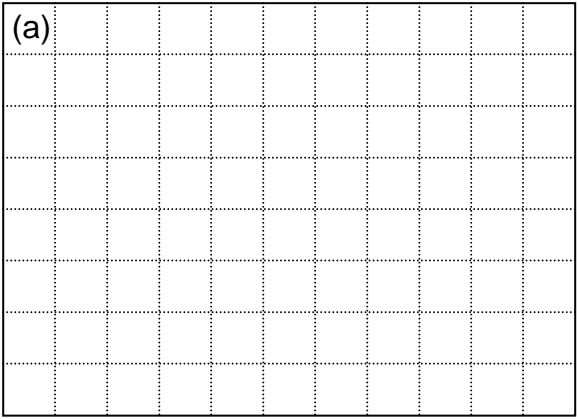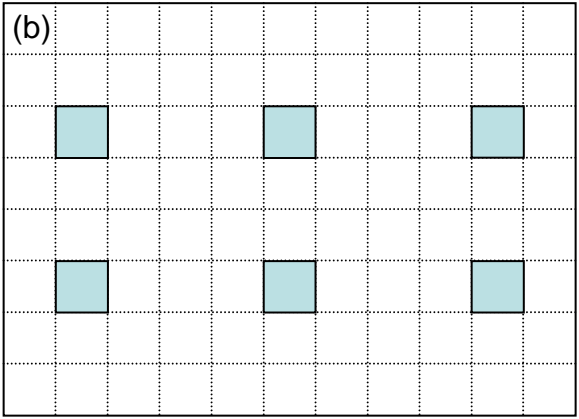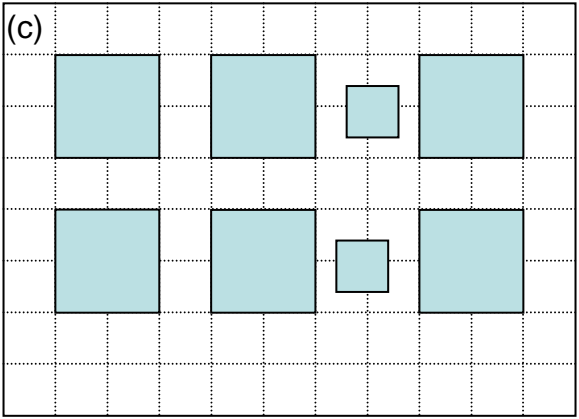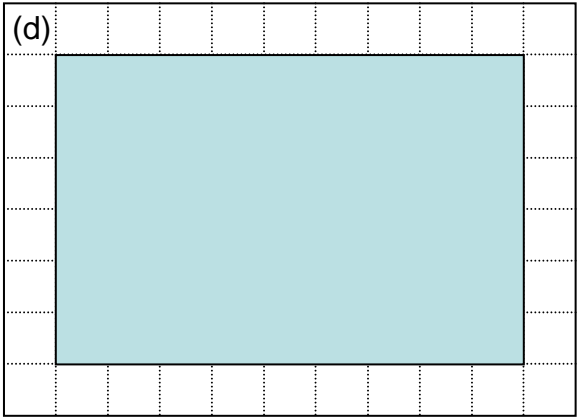

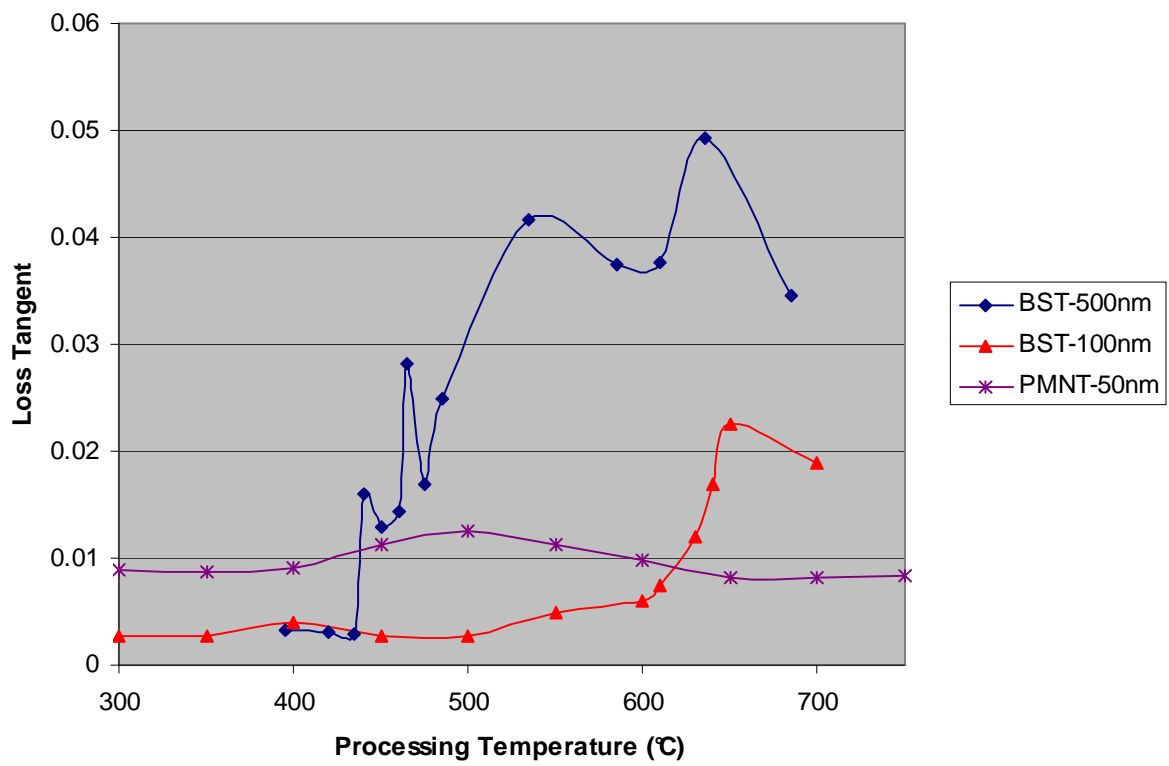

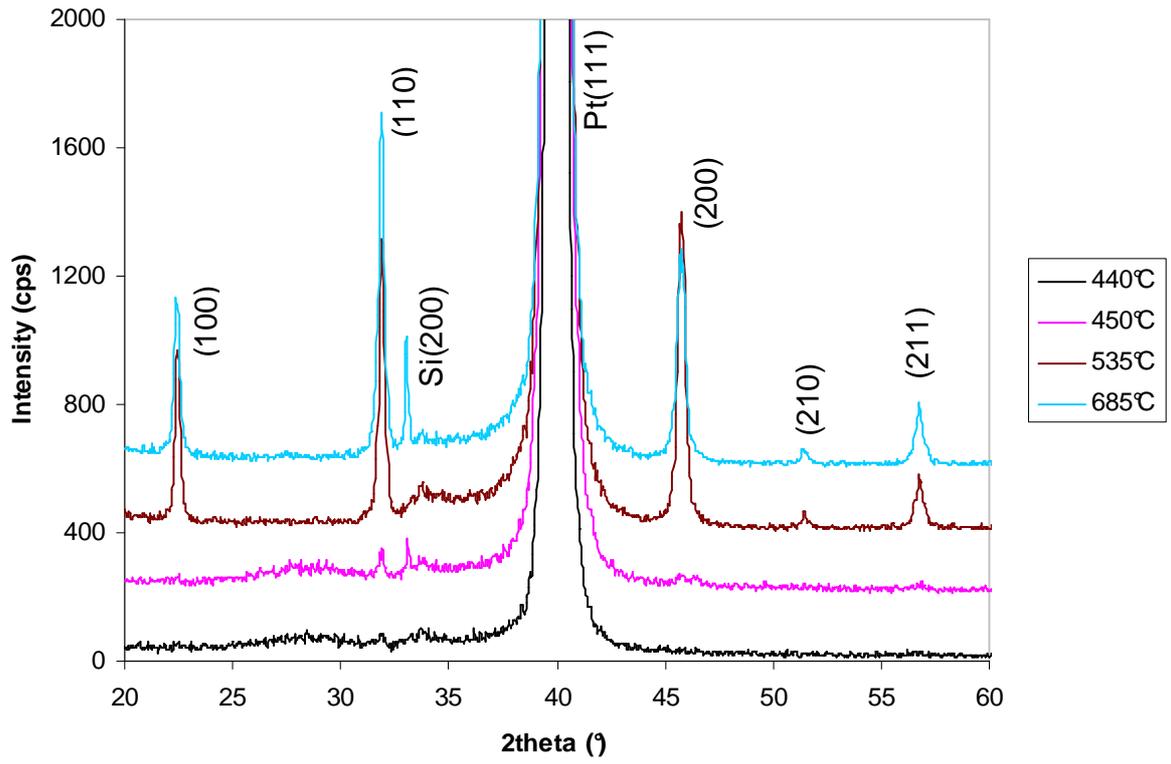

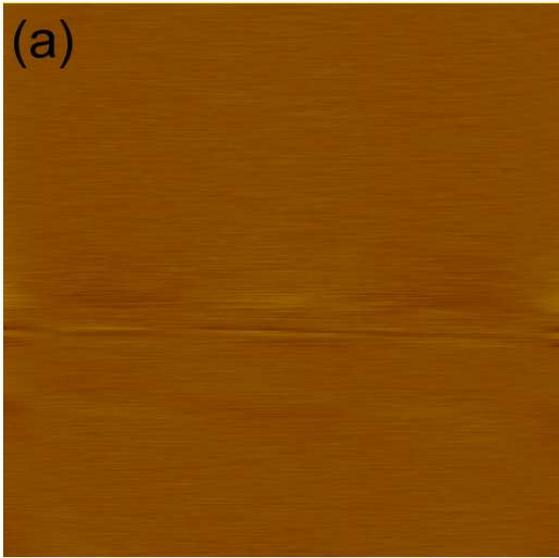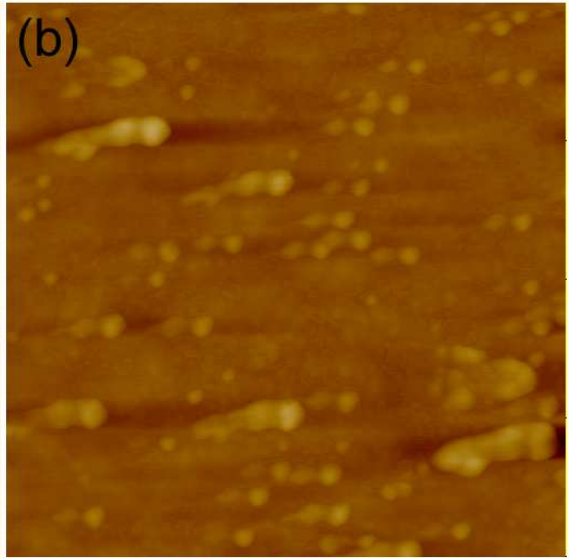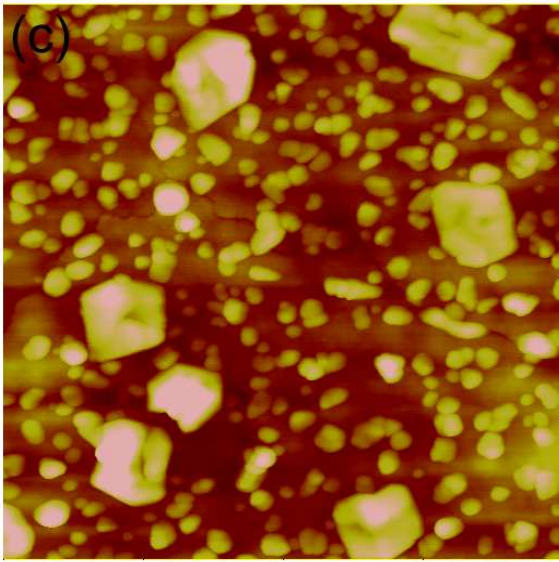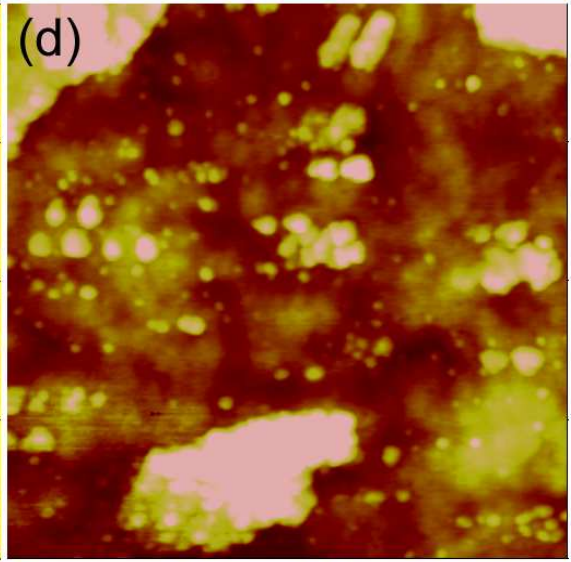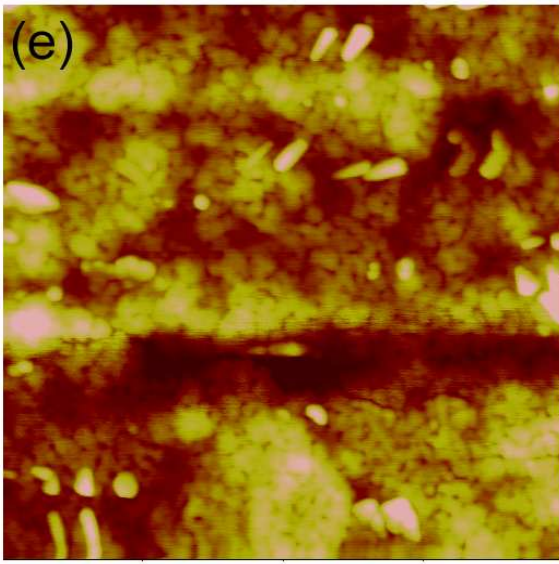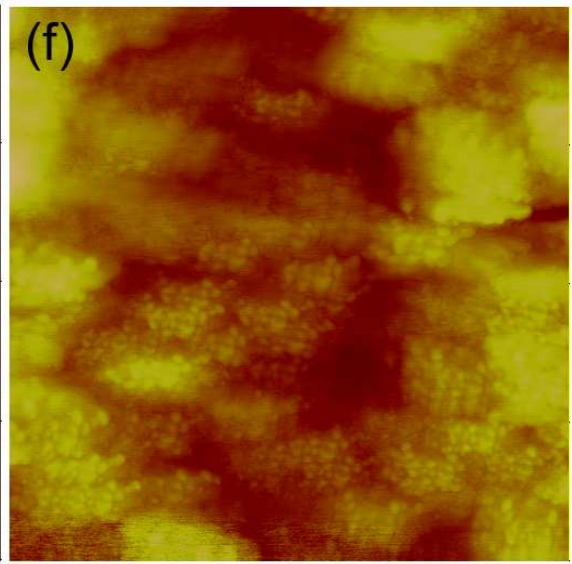

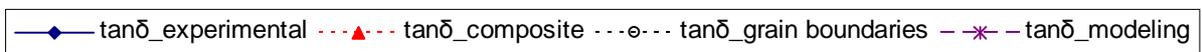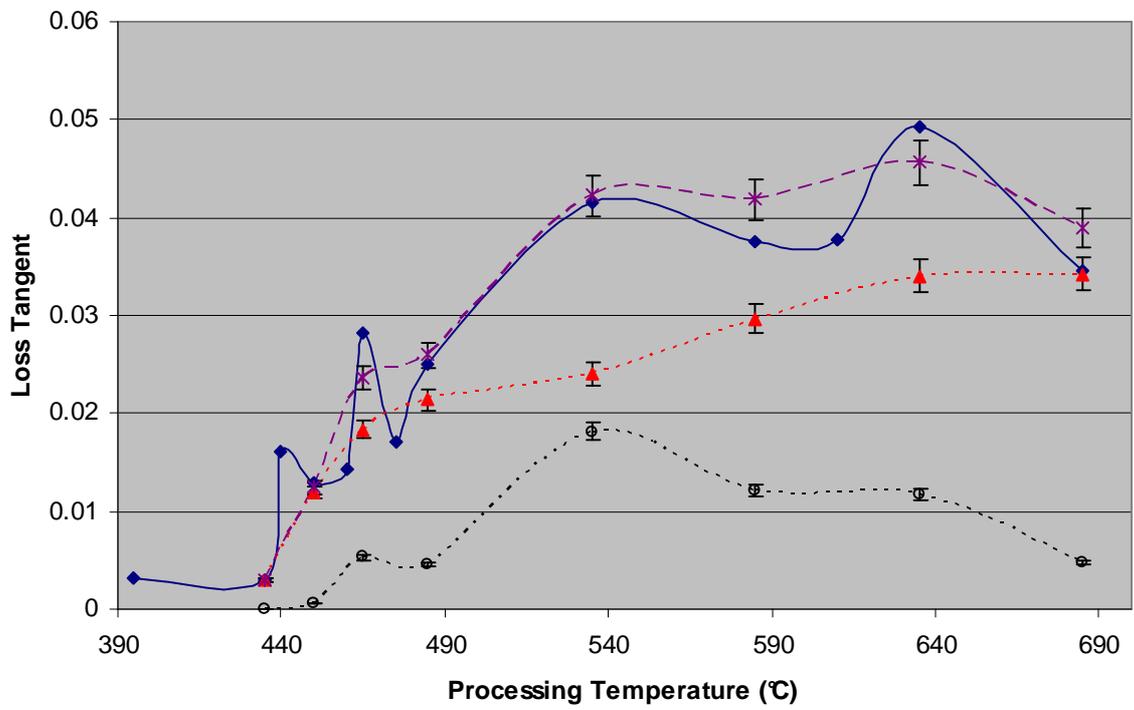